\documentclass[%
 reprint,
 superscriptaddress,
 amsmath,amssymb,
 aps,
]{revtex4-2}

\usepackage[utf8]{inputenc}
\usepackage{graphicx}% Include figure files
\usepackage{dcolumn}% Align table columns on decimal point
\usepackage{bm}% bold math
\usepackage{soul,color,xcolor}             % 用于设置字体颜色
\definecolor{myColor}{rgb}{0,0,0}         % 隐藏修订时用这行
\makeatletter
\newcommand*{\revise}{\@ifnextchar\bgroup{\revise@}{\color{myColor}}}
\newcommand*{\revise@}[1]{{\textcolor{myColor}{#1}}}
\makeatother
\begin{document}

\preprint{APS/123-QED}

\title{Unconventional tunnel magnetoresistance scaling with altermagnets}% Force line breaks with \\

\author{Zongmeng Yang}
\affiliation{State Key Laboratory for Mesoscopic Physics and School of Physics, Peking University, Beijing 100871, P. R. China}
\author{Xingyue Yang}
\affiliation{State Key Laboratory for Mesoscopic Physics and School of Physics, Peking University, Beijing 100871, P. R. China}
\author{Jianhua Wang}
\affiliation{Institute for Superconducting and Electronic Materials, Faculty of Engineering and Information Sciences, University of Wollongong, Wollongong 2500, Australia}
\affiliation{School of Material Science and Engineering, Tiangong University, Tianjin 300387, China}
\author{\revise{Qiang Li}}
\affiliation{\revise{School of Advanced Materials and Mechatronic Engineering, Hubei Minzu University, Enshi 445000, P. R. China}}
\author{Rui Peng}
\affiliation{Science, Mathematics and Technology (SMT) Cluster, Singapore University of Technology and Design, Singapore 487372}
\author{Ching Hua Lee}
\affiliation{Department of Physics, National University of Singapore, Singapore 117551}
\author{Lay Kee Ang}
\affiliation{Science, Mathematics and Technology (SMT) Cluster, Singapore University of Technology and Design, Singapore 487372}
\author{Jing Lu}
\email{jinglu@pku.edu.cn}
\affiliation{State Key Laboratory for Mesoscopic Physics and School of Physics, Peking University, Beijing 100871, P. R. China}
\affiliation{Collaborative Innovation Center of Quantum Matter, Beijing 100871, P. R. China}
\affiliation{Beijing Key Laboratory for Magnetoelectric Materials and Devices (BKL-MEMD), Peking University, Beijing 100871, P. R. China}
\affiliation{Peking University Yangtze Delta Institute of Optoelectronics, Nantong 226010, P. R. China}
\affiliation{Key Laboratory for the Physics and Chemistry of Nanodevices, Peking University, Beijing 100871, P. R. China}
\affiliation{Beijing Key Laboratory of Quantum Devices, Peking University, Beijing 100871, P. R. China}
\author{Yee Sin Ang}
\email{yeesin\_ang@sutd.edu.sg}
\affiliation{Science, Mathematics and Technology (SMT) Cluster, Singapore University of Technology and Design, Singapore 487372}
\author{Shibo Fang}
\email{shibo\_fang@sutd.edu.sg}
\affiliation{Science, Mathematics and Technology (SMT) Cluster, Singapore University of Technology and Design, Singapore 487372}

\begin{abstract}
In conventional magnetic tunnel junctions (MTJs), the tunnel magnetoresistance (TMR) typically increases with barrier thickness as electron transmission in the antiparallel configuration decays faster than that of the parallel configuration. In this work, we reveal an anomalous scaling effect in altermagnetic tunnel junctions (AMTJs), where the TMR decreases anomalously with an increasing barrier thickness. The anomalous scaling originates from the overlapping spin-split branches \revise{forming} a transmission path that cannot be suppressed in the antiparallel state. Such \revise{phenomenon} is explained by a double-barrier model and is further demonstrated using \textit{ab initio} quantum transport simulations in \revise{2D V$_2$Te$_2$O/Cr$_2$Se$_2$O/V$_2$Te$_2$O and V$_2$Te$_2$O/ZnSe/V$_2$Te$_2$O AMTJs}. Our work identifies a peculiar unexpected transport characteristic of AMTJ, providing a fundamental limit on AMTJ device design and illustrating the potential optimal design of AMTJ at the ultrascaled monolayer limit.

\begin{description}
\item[Keywords]
quantum transport simulation, altermagnetic, magnetic tunnel junction, scaling beh-\\ avior, first-principles calculation 
\end{description}
\end{abstract}

\maketitle

%\tableofcontents

\noindent\textit{Introduction.} Magnetic tunnel junctions (MTJs) are a fundamental component of spintronics, which constitute a nonvolatile memory function through the giant magnetoresistance effect~\cite{Han2024, Dieny2020, Schwierz2015, Du2023, Hirohata2020, Zhang2024, Bhatti2017, Zhang2021, WuQing2022}. \revise{The scaling of MTJs is a key trend in the advancement of spintronics to achieve high-density storage, high-speed operation, and efficient spin–orbit torque (SOT)}~\cite{Baltz2018, Jungwirth2016, Han2021, Chen2024, Choi2022, Lee2021, Manchon2019}. Ferromagnetic memory faces difficulties to achieve higher storage density and device miniaturization due to the presence of stray fields~\cite{systems2024, Masuda2024, Han2023, Xiong2022}. Antiferromagnetic tunnel junctions have gained attention for their robustness against magnetic perturbations, absence of stray fields, high N\'eel temperature, and ultrafast spin dynamics~\cite{al2023, Tsymbal2024, Olsen2024, Shao2023, PGu2023, Xchen2023}. However, conventional antiferromagnets exhibit degenerate energy bands, with only hidden spin polarization available for control. Altermagnets (AMs), featuring alternating spin polarizations in both real space and momentum space, have recently been recognized as a type of unconventional magnets~\cite{Tomas2022, Fender2025, Smejkal2022, Liu2025, Kre2024, Song2025, fu_floquet_2025, fu_all-electrically_2025, fukaya_superconducting_2025, ma2021multifunctional, hu2025catalog, han2025nonvolatile, song2025altermagnets, zhou2025manipulation, leihan_discovery_2025}. They exhibit a giant spin-splitting without a net magnetic moment, thus offering an ideal platform for ultracompact integration of MTJs~\cite{Bai2024}.

\begin{figure}[t]
\centering
\includegraphics[width=\linewidth]{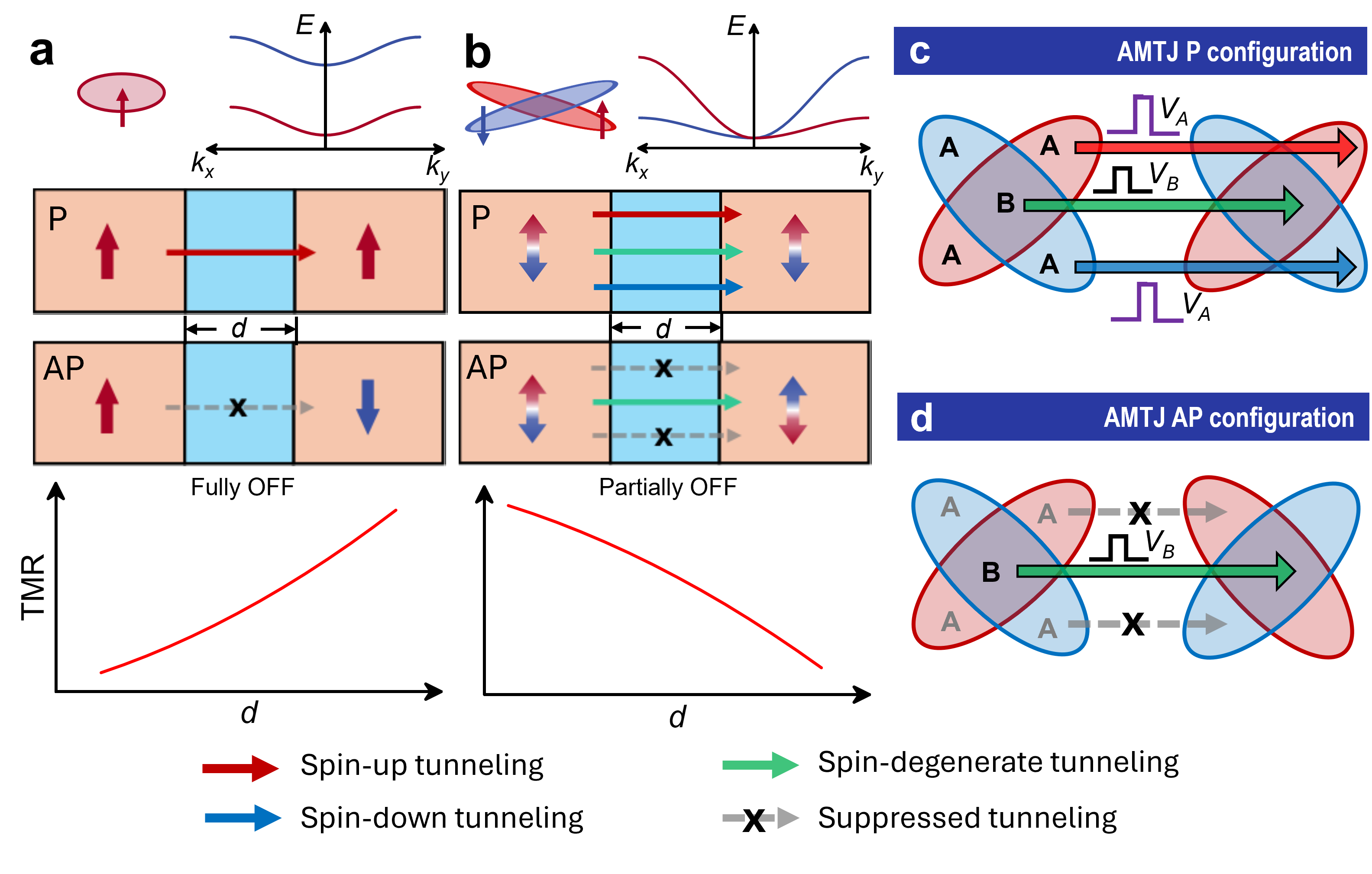} % 自动缩放到半栏宽度
\caption{\label{fig1} Schematic illustration of the tunneling mechanisms in (a) traditional MTJs and (b) AMTJs. The top panels depict the spin-split band structures, while the middle panels show the device configurations 
in the parallel (P) and antiparallel (AP) configurations. The two outer layers represent the magnetic electrodes (either ferromagnetic or altermagnetic), and the central region denotes the insulating barrier 
with thickness $d$. The bottom panels illustrate the dependence of the TMR ratio on $d$, with insets depicting the $k$-space tunneling for electrons in the P and AP configurations. (c) shows the parallel configuration where electronic states in both region A and B contribute to the ON state current. (d) the antiparallel configuration where the transmission of electronic states in region A is suppressed. However, the electronic states in region B continue to facilitate transmission, leading to higher OFF state current.}
\end{figure}

Tunnel magnetoresistance (TMR), a key metric characterizing the performance of MTJs, is defined as
\begin{equation}
\mathrm{TMR} = \frac{G^{\mathrm{P}} - G^{\mathrm{AP}}}{G^{\mathrm{AP}}}
\end{equation}
where \( G^{\mathrm{P}} \) and \( G^{\mathrm{AP}} \) denote the conductance of the device in the parallel (P) and antiparallel (AP) magnetic configurations, respectively ~\cite{Tsymbal2003}. In ferromagnetic and A-type antiferromagnetic tunnel junctions, TMR typically increases as the barrier layer thickness increases~\cite{Yuasa2004, Yang2024, Wu2022, Butler2001, Chi2025, Emoto2024, Zhou2020}. For example, in graphite/2D CrPS$_4$/graphite MTJs, the TMR ratio rises by approximately $10^5$ when the number of CrPS$_4$ layers is increased from 2 to 7~\cite{al2021}. A similar trend is observed in Ag/CrI$_3$/Ag MTJs, where the TMR ratio grows from 10$^3$\% to 10$^9$\% with the barrier thickness of CrI$_3$ varying from 2 to 12 layers~\cite{Wu2022}. This behavior can be explained by the different decay rates of conductance in the P and AP configurations as described by the \textit{Julliere} model~\cite{Butler2001, MacLaren1997, Julliere1975, Tsymbal2003}. As shown in Fig.~\ref{fig1}(a), the conductance is dominated by Bloch states with a small decay rate in P state, which is also called the majority spin conductance. In the AP configuration, the conductance is strongly suppressed due to spin mismatch between the electrodes, and it is mainly contributed by interface resonance states with \revise{a} large decay rate. As a result, TMR decreases as the barrier becomes thinner according to $\mathrm{TMR} \propto \exp(\Delta\gamma_{\mathrm{AP-P}} d)$, where $\Delta\gamma_{\mathrm{AP-P}}$ represents the difference in the decay rate between the AP and the P configurations.

\noindent\textit{Anomalous scaling in AMTJs.} In this work, we reveal an anomalous scaling behavior in altermagnetic tunnel junctions (AMTJs) [Fig.~\ref{fig1}(b)], whose TMR decreases as the barrier thickness increases. This anomalous scaling originates from the distinct morphology of the altermagnetic Fermi surface, where the momentum-space overlap between the spin-split branches forms a persistent transmission channel that cannot be suppressed in the AP configuration. The states that suppress the increase in TMR are, in fact, the spin-degenerate Bloch states in altermagnets. Such spin degeneracy is general in altermagnets and is protected by the symmetry when the little group of momentum includes at least one real-space symmetry operation that connects the opposite-spin sublattices.~\cite{Smejkal2022, Bai2024}. In contrast, ferromagnetic or A-type antiferromagnetic MTJs as discussed above possess fully spin-split bands without overlap, allowing efficient suppression of transmission in the AP configuration. To illustrate the anomalous scaling, we develop a double-barrier model, which \revise{subdivides} the spin-split Fermi surfaces into an outer region and a central region, labeled A and B in Figs.~\ref{fig1}(c) and \ref{fig1}(d), respectively. In the AP configuration, N\'eel vectors of the incident and transmission leads are misaligned. Electron transmission mediated by states in Region A can be effectively switched off due to spin mismatch in this case. However, the transport channel in Region B remains open even in the AP configuration. This arises because of the \revise{overlap} of two spin-split Fermi surfaces which create a spin-degenerate transmission highway that remains open in both P and AP states [Figs.~\ref{fig1}(c) and \ref{fig1}(d)]. As the barrier thickness increases, electron tunneling through the spin-degenerate highway (provided by the electronic states in Region B) remains high with a higher OFF state current. Consequently, TMR reduces with increasing barrier thickness, leading to the anomalous scaling behavior compared to the opposite behavior in conventional MTJs.

To validate this mechanism, we investigate the scaling behavior of AMTJ composed of 2D V$_2$Te$_2$O/Cr$_2$Se$_2$O/V$_2$Te$_2$O using \textit{ab initio} quantum transport simulation. By increasing the Cr$_2$Se$_2$O layer number from 1 to 5 layers, the TMR ratio decreases from 220\% to 40\%, which is in stark contrast to the conventional scaling behavior observed in a wider variety of different MTJs based on different materials ~\cite{Yuasa2004, Yang2024, Wu2022, Butler2001, Chi2025, Emoto2024, Zhou2020}. This work uncovers a previously unrecognized transport mechanism in AMs, emphasizing that spin-degenerate transmission channels can fundamentally limit the spin-selective transport efficiency in AMTJ. \revise{We also show a similar trend appears in the VTO/no-magnetic ZnSe/VTO AMTJ, indicating that such an anomalous scaling effect does not rely on the magnetic properties of the insulating layer.} These findings reveal a potential limitation in the scaling behavior of AMTJ, highlighting the importance of controlling spin-degenerate channels to optimize TMR in AMTJ-based devices.

\noindent\textit{Anomalous scaling mechanism.} According to tunneling theory, the transmission of electrons decays exponentially as $\exp(-2 \kappa d)$ where $d$ is the barrier thickness~\cite{Avishai2013} . Here $\kappa$ is the imaginary part of the wavevector ($k_z = q + i\kappa$), which can be written as
\begin{equation}
\kappa = \sqrt{\frac{2m}{\hbar^2} \left[ V(k_{\parallel}, \sigma) - E_f \right] + k_{\parallel}^2}
\end{equation}
where $k_{\parallel}$ is the wavevector parallel to the tunneling interface, $E_f$ is the Fermi level~\cite{Butler2001, MacLaren1997}, and the potential barrier $V$ is a function of $k_{\parallel}$ and spin $\sigma$. Here, the density of states of the initial and final states are uniform in $k$-space. We denote two different tunneling barriers for spin-degenerate transport mediated by states in Region B and spin-nondegenerate transport mediated by states in Region A as $V_B(k_{\parallel}, \sigma)$ and $V_A(k_{\parallel}, \sigma)$, respectively. In AMTJ, the transport channel associated with $V_B(k_{\parallel}, \sigma)$ is always open regardless of the P and AP configurations. In contrast, the transport channel through $V_A(k_{\parallel}, \sigma)$ can be more effectively switched ON or OFF, thus playing the key role in determining the scaling of TMR. The total conductance under $V_i(k_{\parallel}, \sigma)$ is composed of
\begin{equation}
G_i(k_{\parallel}, \sigma) = G_0 \exp[-2d\kappa_i(k_{\parallel}, \sigma)] \quad (i = A, B)
\end{equation}
where $G_0 = e^2/h$ is quantum conductance ($e$ and $h$ are elementary charge and Planck constant, respectively) and $\kappa_i = \sqrt{2m[V_i(k_{\parallel}, \sigma) - E_f]}/\hbar$, and the index $i = A, B$ corresponds to the contribution of electron transmission mediated through Region A and B, respectively. The total conductance in the P configuration is \revise{$G^\mathrm{P} = \sum_{\sigma, k_{\parallel}\in A}G_i + \sum_{\sigma, k_{\parallel}\in B}G_i$}. In the AP configuration, the tunneling through interface resonance states in Region B can be neglected, resulting in a conductance of \revise{$G^{\mathrm{AP}} = \sum_{\sigma, k_{\parallel}\in B}G_i$}. The TMR ratio is
\revise{
\begin{align}
\label{eq:TMR}
\text{TMR} 
&= \frac{G^\mathrm{P} - G^\mathrm{AP}}{G^\mathrm{AP}} \notag \\
&= \sum_{\sigma} \frac{\sum_{k_{\parallel} \in A} \exp\left[ -2d\kappa_A(k_{\parallel}, \sigma) \right]}{\sum_{k_{\parallel} \in B} \exp\left[ -2d\kappa_B(k_{\parallel}, \sigma) \right]}
\end{align}
}
Here we consider two scenarios: (i) If $\kappa_A<\kappa_B$, the TMR increases with increasing barrier thickness $d$, which is consistent with the conventional scaling behavior. (ii) If $\kappa_A>\kappa_B$, the conductance $G_A$ decays more rapidly than $G_B$ as $d$ increases as shown in Fig.~1(b). As a result, the TMR increases with decreasing $d$, as shown at the bottom of Fig.~1(b), which corresponds to the anomalous scaling effect. While the relative $k_{\parallel}$-space area of regions A and B does not affect the monotonicity of the TMR trend, it will influence the overall decay rate. We next demonstrate this anomalous scaling behavior in a representative AMTJ, corresponding to the second case discussed above.

\begin{figure}[t]
\includegraphics[width=\linewidth]{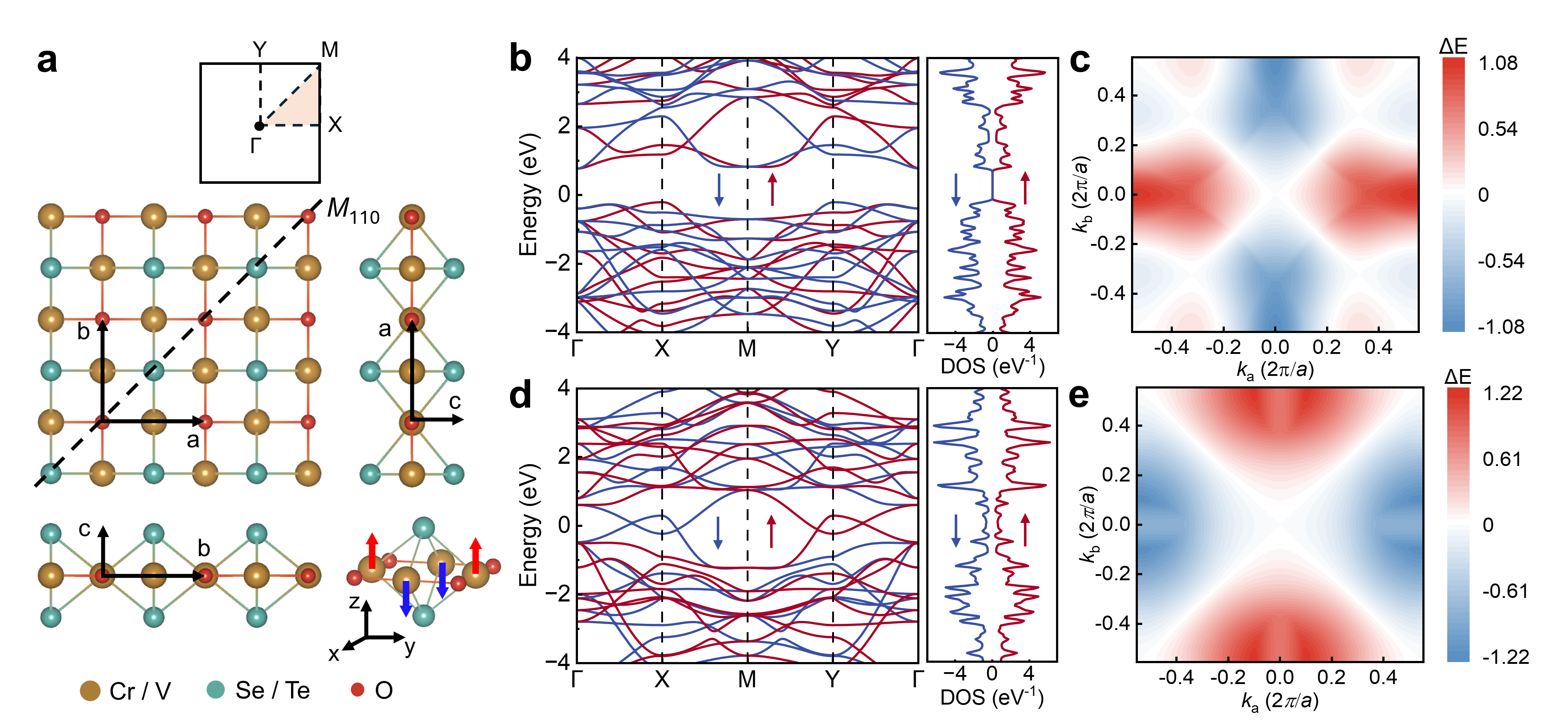}
\caption{\label{fig2}(a) Brillouin zone and crystal structure of Cr$_2$Se$_2$O/V$_2$Te$_2$O. The spin alignment in a unit cell is shown at the bottom-right; band structure and density of states (DOS) for (b) Cr$_2$Se$_2$O, and (c) V$_2$Te$_2$O. The red and blue lines represent the spin-up and spin-down, respectively; spin splitting ($\Delta E = E_{\uparrow} - E_{\downarrow}$) of (d) Cr$_2$Se$_2$O and (e) V$_2$Te$_2$O.}
\end{figure}

\begin{figure}[b]
\centering
\includegraphics[width=\linewidth]{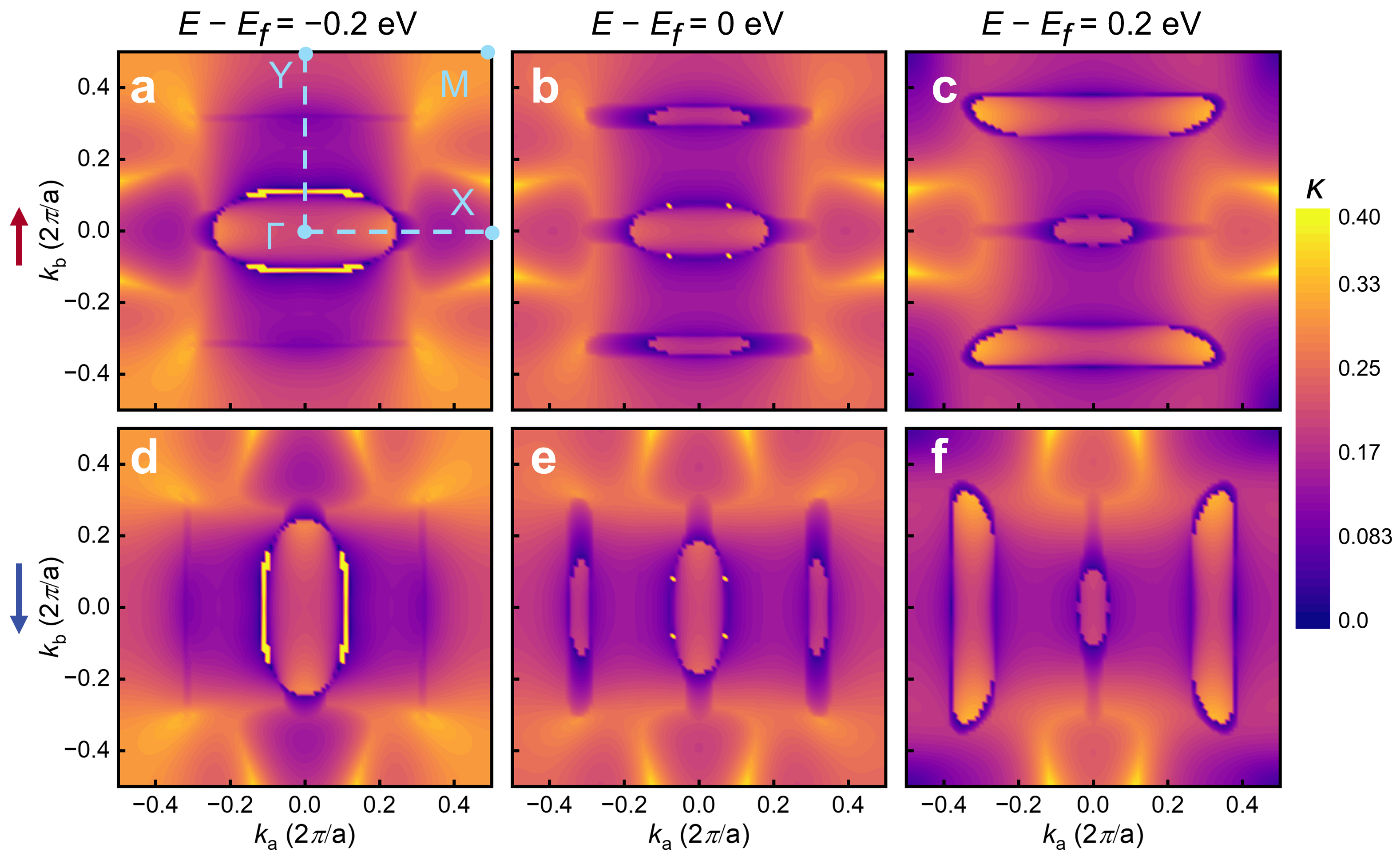} % 自动缩放到半栏宽度
\caption{\label{fig3} Complex band structure in the 2D Brillouin zone, showing the magnitude of $\kappa$. Panels (a), (b), and (c) correspond to spin-up states at $E - E_f = -0.2~\text{eV}$, $E - E_f = 0~\text{eV}$, and $E - E_f = 0.2~\text{eV}$, respectively. Panels (d), (e), and (f) show the corresponding spin-down states of (a), (b), and (c), respectively. The high-symmetry points $\Gamma$, X, Y, and M are labeled in blue.}
\end{figure}

\begin{figure*}
\centering
\includegraphics[width=\linewidth]{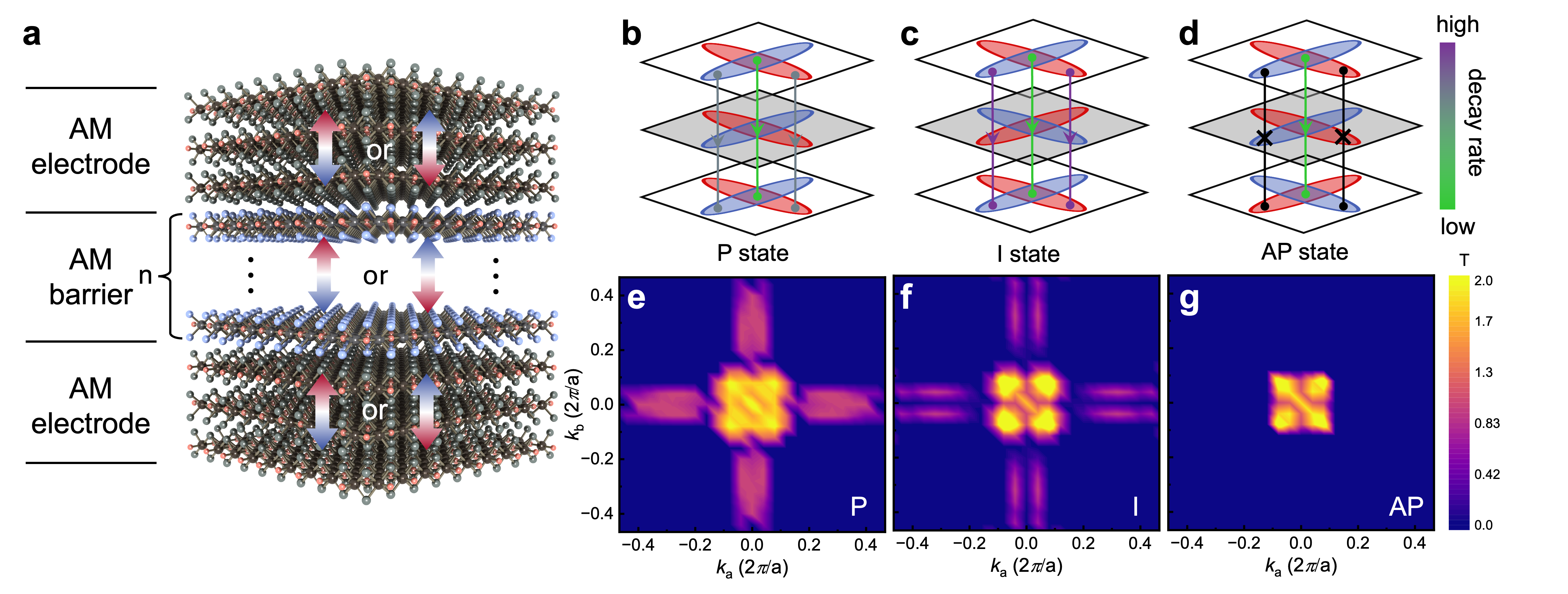} % 自动缩放到半栏宽度
\caption{\label{fig4} (a) Atomic structure of VTO/CSO/VTO AMTJ; $k_{\parallel}$-dependent transmission spectrum at (b) P, (c) I, and (d) AP state; schematic mechanism of tunneling in 2D vdW AMTJ operating at (b) P, (c) I, and (d) AP state, respectively; (e–f) $k_{\parallel}$-dependent transmission spectrum for the configurations corresponding to (a–b).}
\end{figure*}

\noindent\textit{First principle calculation.} We calculate the scaling behavior of 2D AMTJ based on 2D AM metal V$_2$Te$_2$O (VTO) and AM semiconductor Cr$_2$Se$_2$O (CSO). VTO and CSO are both room temperature AM with high N\'eel temperature of 740~K and 350~K, respectively. The lattice mismatch of VTO/CSO heterojunction is less than 0.6\%~\cite{Cui2023, Gong2024}. \revise{The presence of spin-orbital coupling (SOC) can affect both the band structure and spin hybridization of a system. Therefore, we also investigated the impact of SOC on VTO and CSO (see Supplementary Fig. S1~\cite{Material}). The result shows that SOC has a negligible effect on their band structures which is also illustrated in previous studies ~\cite{Cui2023, Gong2024}. Moreover, due to the protection from spin-mirror coupling, the spin hybridization in VTO and CSO is significantly suppressed~\cite{zhang_quantized_2025}. Therefore, SOC is not included in the subsequent calculations.} The atomic structure of these two materials exhibits $M_z$, $M_{110}$, $C_{4z}$, and $C_{2x}$ symmetry [Fig.~2(a)]. The spin polarization of neighboring Cr or V atoms are opposite in the unit cell [bottom right in Fig.~2(a)]. With such magnetic ordering, $PT\tau_{1/2}$ symmetry (where $\tau_{1/2}$ represents a half-lattice translation along [110]) is broken, leading to unconventional spin splitting in reciprocal space. This phenomenon is further validated by the density functional theory (DFT)-calculated monolayer (ML) band structures [Figs.~2(b) and (d)]. The $k$-resolved spin-splitting energy at the conduction band minimum (CBM) of CSO and VTO [Figs.~2(c) and (e)] indicate that they both possess $d$-wave anisotropy with spin degeneration around $\Gamma$ point. This band structure features a spin-degenerate region near the $\Gamma$ point and spin-split regions away from it, corresponding to Region B and Region A in the aforementioned mechanism. The CSO is serving as an insulating barrier with a bandgap of 0.97~eV. The CSO in the insulating layer is AB-stacked, and its structure is shown in Fig.~S3 in Supplementary Material~\cite{Material}. To ensure consistent spin splitting in $k$-space across different layers, an interlayer out-of-plane antiferromagnetic (AFM) spin alignment is adopted within the same material, forming a G-type configuration~\cite{Peng2025}. We construct a tight-binding (TB) model to describe the spin splitting characteristics of the CSO, which is consistent with the results from DFT calculations (Fig.~S2 in Supplementary Material~\cite{Material}). \revise{Nearly identical total energies of both the individual CSO and VTO layers, as well as the CSO/VTO heterostructure under antiferromagnetic (G-type) and ferromagnetic (C-type) configuration, suggest that the original altermagnetic order of each layer is preserved after stacking (Table S1 in Supplementary Material~\cite{Material}).}

The decay $\kappa$ of the tunneling electrons of bulk CSO is obtained by calculating complex band structure~\cite{Mavropoulos2000, Lukashev2012}. Fig.~3 shows the out-of-plane lowest decay rate of spin-up $\kappa_{\uparrow}(k_{\parallel})$ and spin-down $\kappa_{\downarrow}(k_{\parallel})$ electrons of CSO near the Fermi level. The decay characteristics in altermagnets exhibit an alternating nature. Due to spin degeneracy, spin-up and spin-down electrons near the $\Gamma$ point, i.e. in Region A, have nearly identical decay rates. In contrast, Region B corresponds to the spin-split states along the $\Gamma$–X and $\Gamma$–Y directions, where the spin polarization reverses between the two paths with the relation of $\kappa_{\sigma}(\Gamma - X) = \kappa_{-\sigma}(\Gamma - Y)$. The transmitted wave decays more slowly in Region B than in Region A, indicating that the decay channel associated with Region B is more robust than that of Region A.

We consider a VTO/\textit{n}L-CSO/VTO AMTJ composed of an n-layer CSO barrier sandwiched between layered-bulk VTO electrodes (Fig.~\ref{fig4}a). The Néel vectors of both the electrodes and the barrier can reorient independently, giving rise to three distinct magnetic configurations: the parallel (P) configuration, with all Néel vectors aligned (Fig.~\ref{fig4}b); the intermediate (I) configuration, where the barrier is antiparallel to the identically aligned electrodes (Fig.~\ref{fig4}c); and the antiparallel (AP) configuration, where the two electrodes have opposite N\'eel vectors (Fig.~\ref{fig4}d) (see Fig.~S4 for detailed spin alignments for these configurations~\cite{Material}).

We perform DFT calculations coupled with the nonequilibrium Green’s function (NEGF) method, as implemented in QuantumATK (see Supplementary Material for computational details~\cite{Material})~\cite{al2020}, to assess the transport properties of the AMTJ. In the VTO/1L-CSO/VTO AMTJ, switching from the AP state yields TMR ratios of about 220\% for the P state and 137\% for the I state. The difference in transport behaviors between P and I configurations can be attributed to the spin-dependent tunneling barrier. The difference in transmission between the P and I configurations arises from the lower tunneling barrier in the P state, where the spin-splitting of the electrodes and barrier are mutually aligned (Figs.~\ref{fig4}e,f). Likewise, the transmission eigenstates for different spin channels further reveal that the tunneling barrier due to the spin-splitting mismatch between CSO and VTO effectively suppresses electron transmission in the I state (see Figs.~S4 in Supplementary Material~\cite{Material}). The lowest transmission observed in the AP state (Fig.~\ref{fig4}g) indicates that spin matching between electrodes plays a more dominant role in tunneling than spin alignment within the barrier. Especially, spin-degenerate states (near the $\Gamma$ point) consistently exhibit lower tunneling barriers due to the constant spin alignment in all three magnetic configurations and dominate the transport in the AP state.

\begin{figure}[t]
\centering
\includegraphics[width=\linewidth]{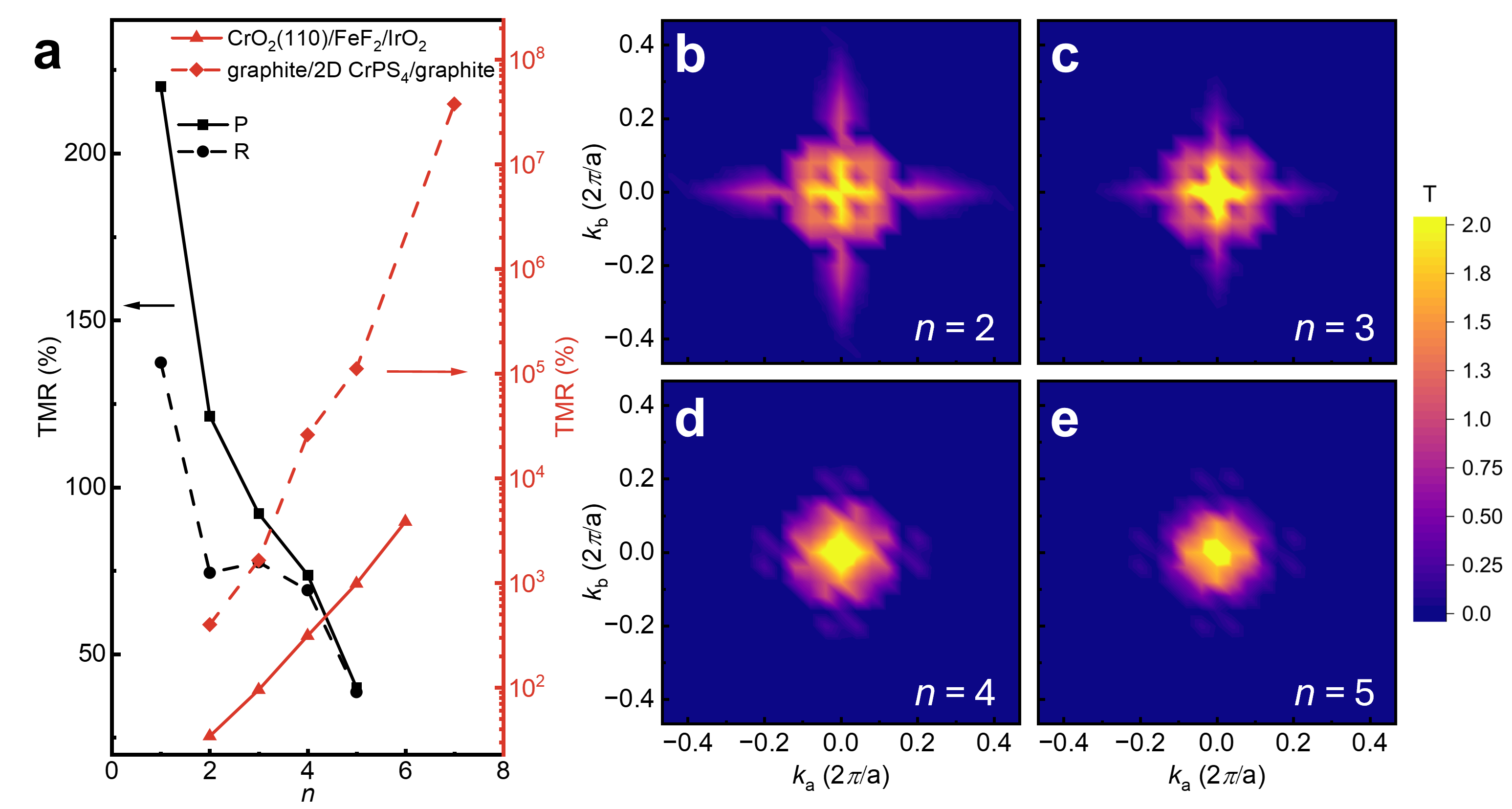} % 自动缩放到半栏宽度
\caption{\label{fig5} (a) TMR as a function of barrier layer number $n$. The black lines represent the TMR of the VTO/$n$-layer CSO/VTO MTJ studied in this work, while the red lines correspond to reference data from Refs.~\cite{Chi2025, al2021}. (b–e) Momentum-resolved transmission spectra at the P state for CSO barrier thicknesses of 2, 3, 4, and 5 layers, respectively.}
\end{figure}

Next, we investigate the tunneling of AMTJ with varying CSO layer number. With increasing CSO thickness, all three magnetic configurations exhibit exponential decay in transmission. However, the decay rates differ significantly: the AP configuration shows the slowest decay (only 31\%), in contrast to 72.7\% and 59.3\% reductions in the P and I states, respectively. This trend highlights a persistent transmission channel in the AP configuration that is less sensitive to barrier thickness. As a result, the TMR drops anomalously: from 220\% to 40\% and from 137\% to 39\% for the P-to-AP and I-to-AP switching, respectively, as the number of CSO layers increases from 1 to 5 (Fig.~\ref{fig5}a). \revise{Theoretically, TMR continues to decay with increasing barrier thickness and asymptotically approaches zero. In practice, however, as the tunneling current diminishes with thickness, it may eventually fall below the detection limit of experimental setups. These considerations imply two distinct critical thicknesses: one defined by the experimental measurability of the current, and the other by the theoretical vanishing of TMR. The relative importance of each depends on the specific material parameters and measurement conditions.} The detailed energy-dependent transmissions for different magnetic configurations are shown in Fig.~S4(b)–(c) (Supplementary Material~\cite{Material}). 

\begin{figure}[b]
\centering
\includegraphics[width=\linewidth]{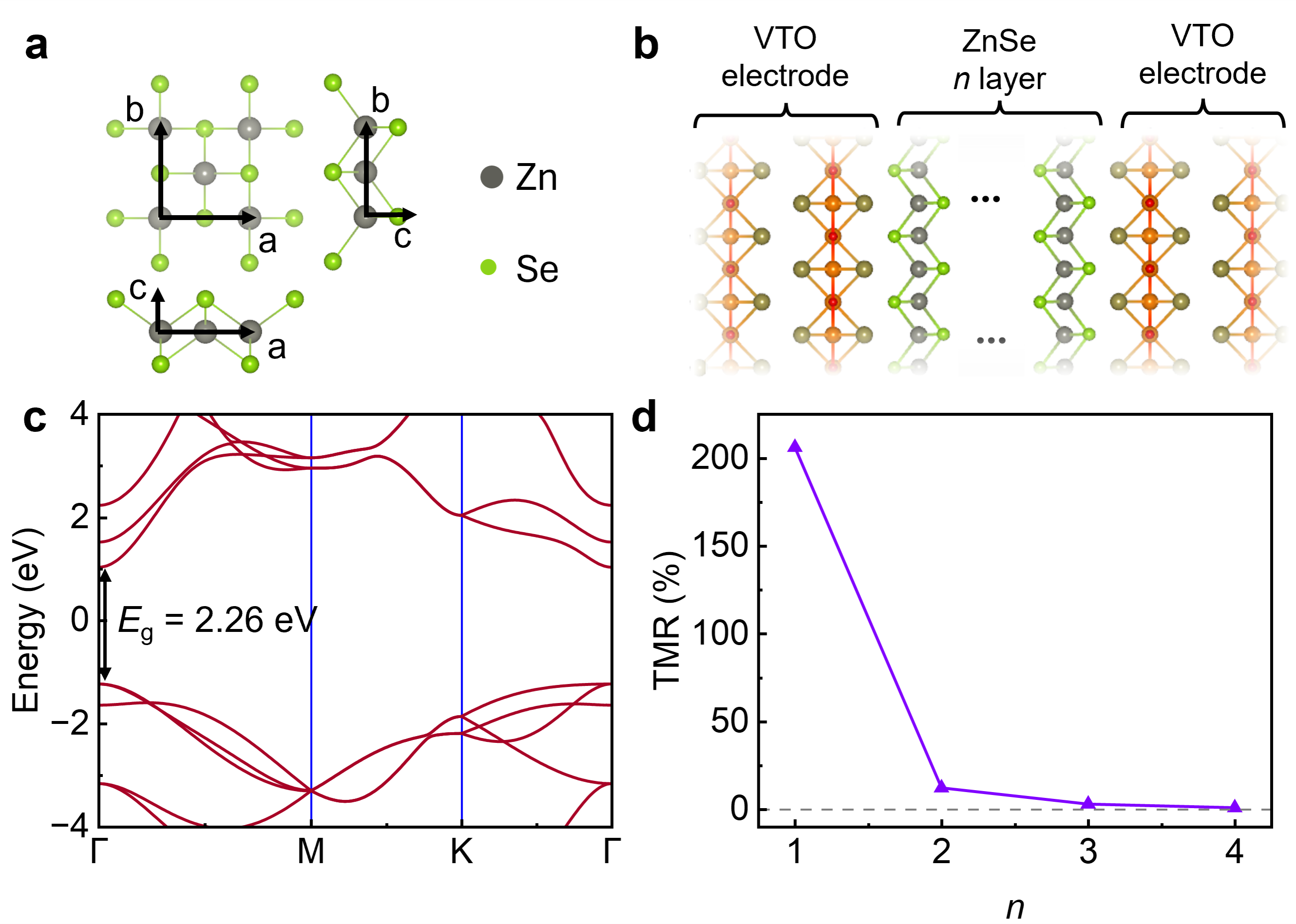} 
\revise{\caption{\label{fig6} (a) Crystal structure of ZnSe; (b) Atomic structure of the VTO/n-layer ZnSe/VTO AMTJ; (c) Bandstructure of monolayer ZnSe; (d) TMR of the VTO/n-layer ZnSe/VTO AMTJ as a function of ZnSe thickness}}
\end{figure}

\revise{\noindent\textit{Discussion and conclusion.} Since all-altermagnetic MTJ geometry may suffer from magnetic instability due to strong exchange interaction between adjacent altermagnetic layers~\cite{klein_enhancement_2019}, we adopt ZnSe as a non-magnetic barrier and VTO as electrodes[Figs.~\ref{fig6}(a)(b)]. ZnSe is a semiconductor with a direct bandgap of 2.26 eV as shown in Figs.~\ref{fig6}(c) ~\cite{xuan_ferroelasticity_2022}, and its complex bandstructure reveals a relatively small decay rate near the $\Gamma$ point [Fig. S6(a) in Supplementary Material]. In the VTO/ZnSe/VTO AMTJ, the TMR decreases from 206\% to 0.9\% with increasing barrier thickness [Fig.~\ref{fig6}(d)], exhibiting a trend similar to that in the above all-altermagnetic case. This behavior originates from the coexistence of spin-degenerate and spin-nondegenerate Bloch states in altermagnets, which possess distinct decay rates through the barrier. When the barrier selectively favors the more slowly decaying spin-degenerate channel, the unconventional scaling emerges, regardless of whether the barrier itself is altermagnetic. Detailed transmission spectra are provided in Fig. S6 of the Supplementary Material.}

\revise{In addition, the applicability of our theoretical framework may potentially be extended to a wide range of momentum–spin coupled systems, such as spin–orbit coupling (SOC)~\cite{huang_two-dimensional_2025, tao_ferroelectric_2025}, noncollinear antiferromagnets~\cite{rimmler_non-collinear_2025}, and spin–valley locking~\cite{yan_giant_2025}, etc. The key point of this effect lies in the fact that spin-degenerate and non-degenerate Bloch waves exhibit different decay rates. In addition, pinholes or resonant tunneling in conventional antiferromagnetic devices may also lead to anomalous scaling behavior~\cite{suzuki_tunnel_2023}. This anomalous scaling behavior in TMR does not necessarily occur only in altermagnets and therefore cannot be used as definitive evidence for identifying altermagnetic materials.}

In summary, we reveal an anomalous scaling behavior of decreasing TMR with increasing barrier thickness in AMTJ, which deviates from conventional expectations. The underlying mechanism of this anomalous scaling lies in the presence of a persistent transmission channel arising from spin-degenerate states, which remains largely unaffected by switching between the P and AP configurations. To qualitatively interpret this behavior, we further construct a simplified double-barrier model that distinguishes between effective and ineffective spin-selective transmission. The effective part, originating from spin-split states, decays rapidly with increasing barrier thickness. While the ineffective part exhibits weak decay and ultimately dominates the overall transmission at larger thicknesses. Whether the TMR exhibits an unconventional scaling behavior is primarily determined by the difference in decay constants $\kappa$ between the ineffective and effective regions. The overall decay rate of TMR with barrier thickness is further influenced by both the relative area ratio of these $k$-space regions and their respective decay rates. \revise{To confirm this unconventional behavior, we perform \textit{ab initio} quantum transport simulations on all-altermagnetic V$_2$Te$_2$O/Cr$_2$Se$_2$O/V$_2$Te$_2$O  AMTJs and AMTJs with a non-magnetic ZnSe barrier.} Our results not only provide new insights into the fundamental transport behavior of altermagnets but also highlight a key design consideration that fundamentally impacts the performance of the magnetic tunnel junction based on altermagnets.

\noindent\textit{Experimental implementation.} We note that bulk materials with similar structure to these two materials, KV$_2$Se$_2$O and SbV$_2$Te$_2$O, have been successfully synthesized in recent experiments~\cite{al2025, FZhang2025}. Spin- and angle-resolved photoemission spectroscopy (SARPES) measurements confirm them as $d$-wave anisotropic AM. Moreover, its 2D form can be obtained via topochemical deintercalation~\cite{Lin2018}, thus suggesting V$_2$Te$_2$O and Cr$_2$Se$_2$O as potential feasible material platform for experimental realization.

\begin{acknowledgments}
This work was supported by the National Natural Science Foundation of China (No.~12274002 and 91964101), the Ministry of Science and Technology of China (No.~2022YFA1203904), the China Scholarship Council (CSC), the Fundamental Research Funds for the Central Universities, the High-performance Computing Platform of Peking University and the National Research Foundation, Singapore, under its Frontier Competitive Research Programme (NRF-F-CRP-2024-0001). L.~K.~A.\ is supported by the Singapore ASTAR IRG (M23M6c0102). We thank Dingfu Shao for insightful discussions during 2025 Workshop on Altermagnets and School on Magnetic Symmetry (Shanghai). The authors also thank Dr. Yuntian Liu for helpful discussions.
\end{acknowledgments}

\bibliography{apssamp}

\end{document}